\documentclass[twocolumn,superscriptaddress,amsmath,showpacs,showkeys,aps,pre,eqsecnum]{revtex4-1}
\usepackage[utf8]{inputenc}
\usepackage[final]{graphicx}
\usepackage{bm}
\usepackage[normalem]{ulem}
\usepackage[usenames]{color}
\usepackage{times}
\usepackage{verbatim}
\usepackage{xcolor}
\usepackage{hyperref}
\usepackage{wrapfig}
\usepackage{subfigure}
\DeclareGraphicsExtensions{.eps}
\usepackage{upgreek}
\usepackage{mathrsfs} 

\begin{document}
\title{State dependent diffusion in a bistable potential: conditional probabilities and escape rates}
\author{Miguel V. Moreno}
\affiliation{Instituto de de F\'\i sica Te\'orica, Universidade Estadual Paulista, Rua Dr. Bento Teobaldo Ferraz 271,  01140-070 
S\~ao Paulo, SP, Brazil.}
\altaffiliation{Previously at:  Departamento  de  F\'\i sica, Universidade  Federal  Fluminense   and  National  Institute  of  Science  and  
Technology  for Complex Systems, Av.  Gal.  Milton Tavares de Souza s/n, Campus da Praia Vermelha, 24210-346 Niter\'oi, RJ, Brazil }

\author{Daniel G. Barci}
\affiliation{Departamento de F{\'\i}sica Te\'orica,
Universidade do Estado do Rio de Janeiro, Rua S\~ao Francisco Xavier 524, 20550-013,  Rio de Janeiro, RJ, Brazil.}
\author{Zochil Gonz\'alez Arenas}
\affiliation{Departamento de Matem\'atica Aplicada, IME, Universidade do Estado do Rio de Janeiro, Rua S\~ao Francisco 
Xavier 524, 20550-013,  Rio de Janeiro, RJ, Brazil}

\date{\today}

\begin{abstract}
We consider a simple model of a bistable system under the influence of multiplicative noise. We provide a path integral representation 
of the overdamped Langevin dynamics and compute conditional probabilities and escape rates in the weak noise approximation. 
The saddle-point solution of the functional integral is given by a diluted gas of instantons and anti-instantons, similarly to the additive 
noise problem. However, in this case, the integration over fluctuations is more involved. We  introduce a local time reparametrization 
that allows its computation in the form of usual Gaussian integrals. We found corrections to the Kramers' escape rate produced by the 
diffusion function which governs the state dependent diffusion for arbitrary values of the  stochastic prescription parameter. Theoretical 
results are confirmed through numerical simulations.
\end{abstract}

\maketitle

\section{Introduction}
The physics of thermal or noise activation over a barrier has a long history. Nowadays, it is an important  research topic due to 
the wide range of  applications in several areas of science, such as physics, chemistry and biology as well~\cite{ActivatedBarrierBook}. 
The simplest model to  study this problem is a classical particle in a bistable potential, $U(x)$, whose dynamics is driven by an overdamped 
Langevin equation with  additive  white noise. In this context, an important physical quantity is the rate at which the particle escape
 out of a minimum of the potential. 
The seminal work of Kramers~\cite{Kramers1940} stated the very simple formula 
\begin{equation}
r_{\rm add}= \frac{\sqrt{\omega_{\rm min}|\omega_{\rm max}|} }{2\pi} e^{-\frac{\Delta U}{\sigma^2}}
\label{eq:KramersEscapeRate}
\end{equation}  
 where $r_{\rm add}$ is the escape rate, $\Delta U=U(x_{\rm max})-U(x_{\rm min})$ is the height of the potential barrier, 
 $\sigma^2$ is the noise intensity and $\omega_{\rm min}=U''(x_{\rm min})$ and $\omega_{\rm max}=U''(x_{\rm max})$ are the local curvatures 
 of the potential at its minimum ($x_{\rm min}$) and its maximum ($x_{\rm max}$), respectively (primes mean derivative with respect to $x$). 
 We use the notation $r_{\rm add}$ to emphasize that this expression for the  escape rate was computed assuming an {\em additive noise} stochastic 
 differential equation.
 Equation~(\ref{eq:KramersEscapeRate}) is valid in the weak noise or high barrier approximation $\sigma^2\ll \Delta U$. 
 Since this well-established result was defined, a lot of work has been done in order to compute more accurate expressions suitable to be applied 
 to more realistic situations. The  generalization of Eq.~(\ref{eq:KramersEscapeRate})  to multidimensional systems was (and still is) a big 
 challenge~\cite{Hangii1990}. Moreover, generalizations to different types of noise probability distributions have been also 
 considered~\cite{BrayMcKane1989,McKane1-1990,McKane2-1990,McKane3-1990,Jung2005,Goulding2007}.

 On the other hand, there is an increasing interest for multiplicative noise stochastic systems. Some examples of multiplicative noise 
 dynamics are given by  the diffusion of particles near a wall~\cite{Lancon2001,Lancon2002,Lubensky2007,Volpe2010,Volpe2011}, 
 micromagnetic dynamics~\cite{GarciaPalacios1998, Aron2014, Arenas2018} and non-equilibrium transitions into absorbing states~\cite{Hinrichsen2000}. 
 There are two particular stochastic phenomena in which multiplicative noise plays an important role: noise-induced phase 
 transitions~\cite{Parrondo1994,CastroWio1995,Sancho2003,Goldenfeld2015,BarciMiguelZochil2016} and  stochastic 
 resonance~\cite{Benzi1981,Parisi1983,Wio2007,Wio2002}. In the last case, the escape rate is at the stem of the physical description of 
 the observed phenomenology.  

 One of the main questions that we address in this paper is how the Kramers' escape rate of Eq.~(\ref{eq:KramersEscapeRate}) is modified 
 when the dynamics is driven by  a general  multiplicative noise, modeled by a diffusion function $g(x)$.  This topic have been rarely treated 
 in the past and there is some controversy in the literature~\cite{Jin2005, FengGuo2011,NingLi2006,Zheng2011, Rosas2016}. 
 In particular, we study  the dependence of the escape rate on the stochastic prescription, necessary to correctly define the multiplicative noise 
 Langevin equation. This point is particularly relevant in order to compare  analytic results with numerical simulations.   
 Our main result is
 \begin{align} 
  r_{\rm mult}=g^2(x_{\rm max})  \frac{ \sqrt{\tilde\omega_{\rm min}|\tilde \omega_{\rm max}| }}{2\pi}\; e^{-\frac{\Delta U_{\rm eq}}{\sigma^2}}\; . 
  \label{eq:mainresult}
 \end{align} 
 We used the notation $r_{\rm mult}$ to denote the escape rate in the {\em multiplicative noise} case. In general, we observe that the 
 Arrhenius form of  the Kramers' result still remains. Another similarity with Eq.~(\ref{eq:KramersEscapeRate}) is that the escape rate does 
 not depend on details, either of the potential or of the diffusion function. Instead, it only depends on the local properties of these 
 functions at the maximum and minima of the potential. On the other hand, there are significant differences between both results. Firstly, 
 the original potential $U(x)$ has been replaced by the equilibrium potential $U_{\rm eq}(x)$, obtained from the solution of the asymptotic 
 stationary Fokker-Planck equation (Eq.~(\ref{eq:Ueq})). This potential depends on the noise and, more important, on the prescription used to 
 interpret the stochastic differential equation.  
 The barrier height is, in this case, $ \Delta U_{\rm eq}=U_{\rm eq}(x_{\rm max}) - U_{\rm eq}(x_{\rm min})$. 
 It is worth to note that $x_{\rm max}$ and $x_{\rm min}$ are the position of the maximum and minumum of the equilibrium potential 
 $U_{\rm eq}$ and not of the original ``classical'' potential $U(x)$. 
  Local curvatures  $\tilde\omega_{\rm min}=U''_{\rm eq}(x_{\rm min})$ and 
 $\tilde\omega_{\rm max}=U''_{\rm eq}(x_{\rm max})$ are also computed by using  the  equilibrium potential.  
 Finally, there is an overall factor given by the diffusion function computed at the maximum of the equilibrium potential, 
 $g^2(x_{\rm max})$,  coming from a 
 careful treatment of fluctuations.  We  describe the model and the technique used to compute 
 Eq.~(\ref{eq:mainresult}), discussing the result in more detail, throughout the paper. 
 
 Multiplicative stochastic processes can be studied with different theoretical approaches. For numerical simulations~\cite{Sivak2013}, the 
 Langevin approach seems to be more adequate. The Fokker-Planck equation is perhaps more appropriate to develop analytic calculations, 
 specially in the long time stationary limit. In this context, techniques  such as mean fields, perturbation theory and even 
 renormalization group  are also available~\cite{Goldenfeld}.  On the other hand,  the path integral formulation of stochastic processes 
 is the more natural technique  to compute correlation and response functions~\cite{WioBook2013}. 
 Important progress has been recently reached in the path integral representation of multiplicative noise 
 processes~\cite{AronLeticia2010, arenas2010, arenas2012, Arenas2012-2, Miguel2015, ArBaCuZoGus2016}, despite the fact that this topic 
 has been studied for a long time~\cite{Janssen-RG}.

 The escape rate is just one ingredient of a more general problem that is the computation of conditional probabilities. 
 Equilibrium properties, such as detailed balance,  can be cast in terms of the conditional probability and its time reversal. 
 Time reversal transformations, detailed-balance relations, as well as microscopic reversibility in multiplicative processes were studied in 
 detail  in Ref.~\cite{Arenas2012-2}. More recently, we have presented a useful path integral technique to compute weak noise 
 expansions~\cite{Miguel2019}. The integration over fluctuations in the multiplicative case is not trivial. The reason is that the 
 diffusion function produces an integration measure that resembles a curved time axis~\cite{Zinn-Justin}. We have provided a local time 
 reparametrization  in order to integrate fluctuations~\cite{Miguel2019}.  In this paper, we compute  the conditional probability of finding a 
 particle in a well at large times $t/2$, provided it was in the same or the other well  at $-t/2$. In the weak noise approximation, 
 saddle points  provide a set of diluted  instanton and anti-instanton solutions. The diluted instanton gas approximation was first introduced 
 in the context of quantum mechanics to compute the tunneling probability across a potential barrier~\cite{Coleman1979}. In the context of 
 an additive stochastic process, it  was developed with great detail in Refs.~\cite{Caroli1979,Caroli1981}.  From a technical point o view, 
 we generalize  the calculation of Ref.~\cite{Caroli1981} to the multiplicative noise case, using the time reparametrization techniques 
 introduced in Ref.~\cite{Miguel2019}. We also perform extensive Langevin simulations to test our results and approximations, 
 finding an excellent agreement. 

 The paper is organized as follows. In the next  section, we present the equilibrium properties of a particle in a double-well 
 potential under state dependent diffusion. In section~\ref{sec:pathintegral}, we briefly review the path integral representation 
 of a conditional probability in a multiplicative process and we show, in section~\ref{sec:Weaknoise}, how to integrate fluctuations.  
 We develop the dilute instanton  gas  approximation in section~\ref{sec:instanton}, where we compute conditional probabilities and 
 the escape rate. In~\ref{Ap:Numerics} we present Langevin simulations of a  particular model and compare the output with our analytic 
 results. Finally, we discuss our results in section~\ref{sec:discussion}. We lead to the Appendix~\ref{Ap:Zeromode} some 
 details of the calculation. 
 
 \section{Equilibrium properties of a particle in a double-well potential under state dependent diffusion}
 \label{sec:model}
 In this section, we describe the equilibrium properties of a  model consisting of a single  particle in a  double-well potential 
 coupled with a thermal bath with state dependent diffusion. 
 We consider a conservative one dimensional system described by a potential energy $U(x)=U(-x)$ with a double minima structure. 
 The thermal bath is characterized by the diffusion function $g(x)=g(-x)$.  The reflection symmetry $x\to -x$ is not essential and most 
 of our results do not depend on it. However, to keep the discussion as simple as possible, we focus  in the  symmetric model,  leading 
 the details of a more general asymmetric  situation to a future presentation.

 In order to reach thermodynamic equilibrium at long times, the drift force $f(x)$ should be related with the classical potential $U(x)$ 
 through a generalized Einstein relation~\cite{arenas2012, Arenas2012-2}
 \begin{equation}
  f(x)=-\frac{1}{2} g^2(x) \frac{dU(x)}{dx}\;. 
  \label{eq:Einstein}
 \end{equation}
 In this way,  the overdamped stochastic dynamics is driven by the Langevin equation
 \begin{equation}
  \frac{dx}{dt} =-\frac{1}{2} g^2(x) \frac{dU(x)}{dx}  + g(x) \eta(t),
  \label{eq:Langevin2} 
 \end{equation}
 where $\eta(t)$ obeys a Gaussian white noise distribution with
 \begin{equation}
  \left\langle \eta(t)\right\rangle   = 0 \;\;\mbox{,}\;\;\;  \left\langle \eta(t)\eta(t')\right\rangle = \sigma^2 \delta(t-t')\; , 
  \label{eq:whitenoise}
 \end{equation}
in which $\sigma^2$ measures the noise intensity.
This equation is  understood in the  
{\em generalized Stratonovich}~\cite{Hanggi1978} prescription 
(also known as $\alpha-$prescription~\cite{Janssen-RG}).
The asymptotic long time equilibrium probability distribution  is given by~\cite{Arenas2012-2}
\begin{equation}
P_{\rm eq}(x)={\cal N}\; e^{-\frac{1}{\sigma^2}U_{\rm eq}(x)},
\label{eq:Peq}
\end{equation} 
where ${\cal N}$ is a normalization constant and the equilibrium potential 
\begin{equation}
U_{\rm eq}(x)= U(x)+(1-\alpha)\sigma^2\ln g^2(x)\; .
\label{eq:Ueq}
\end{equation} 
The parameter $0\le \alpha\leq 1$ labels the particular stochastic prescription used to discretized the Langevin equation. For instance, 
$\alpha=0$ corresponds with It\^o interpretation while $\alpha=1/2$ corresponds with the Stratonovich one.  
In this way, the equilibrium potential is not the bare classical potential, but it is corrected by the  diffusion function $g(x)$. 
On the other hand, the case  $\alpha=1$ corresponds with H\"anggi-Klimontovich or kinetic interpretation~\cite{Hanggi1982,Klimontovich}. 
This is the only prescription which leads to the Boltzmann distribution $U_{\rm eq}(x)=U(x)$. 
 Furthermore, this prescription is also known as 
anti-It\^o   and can be considered as the time reversal conjugated to 
the It\^o prescription~\cite{Arenas2012-2,Miguel2015}.

Although the techniques and results of this paper do not depend on details, either of  $U(x)$ or of $g(x)$, it is convenient,  
just to visualize  the equilibrium potential $U_{\rm eq}(x)$, to  consider a very simple model. Let us take, for instance,  
\begin{equation}
U(x)=-\frac{1}{2} x^2 +\frac{1}{4} x^4 \; , 
\label{eq:U}
\end{equation}
with the diffusion function 
\begin{equation}
g(x)=1+\lambda x^2.
\label{eq:g}
\end{equation}
where the parameter $\lambda$ measures in some sense the multiplicative character of the noise. The particular value of $\lambda=0$ 
corresponds with an additive noise. The potential $U(x)$ has two degenerated minima at $x_{\rm min}=\pm 1$ and a local maximum at $x_{\rm max}=0$.
The contribution of the multiplicative noise for the equilibrium potential is quite interesting. In the weak noise limit, the 
global two-minima structure remains the same.
However, the minima are displaced to   
\begin{align}
x_{\rm min}&=\pm (1-4\sigma^2(1-\alpha))^{1/4}\nonumber \\
& \sim \pm 1\mp\sigma^2(1-\alpha)+ O(\sigma^4)\; .
\end{align}
For $\sigma^2\ge 1/4(1-\alpha)$, both minima melt in a single one,  deeply changing the global structure of the potential. This dependence 
on the noise intensity resembles a second order phase transition, where the critical noise  is given by
\begin{equation}
\sigma_c=\frac{1}{2}\frac{1}{\sqrt{1-\alpha}}\;.
\end{equation}
Interestingly, the critical noise depends on the stochastic prescription. For  $\alpha\to 1$, $\sigma_c\to \infty$, meaning that, 
in the anti-It\^o prescription, the double-well structure is preserved for all values of the noise. 
\begin{figure}
\includegraphics[height= 4.7 cm]{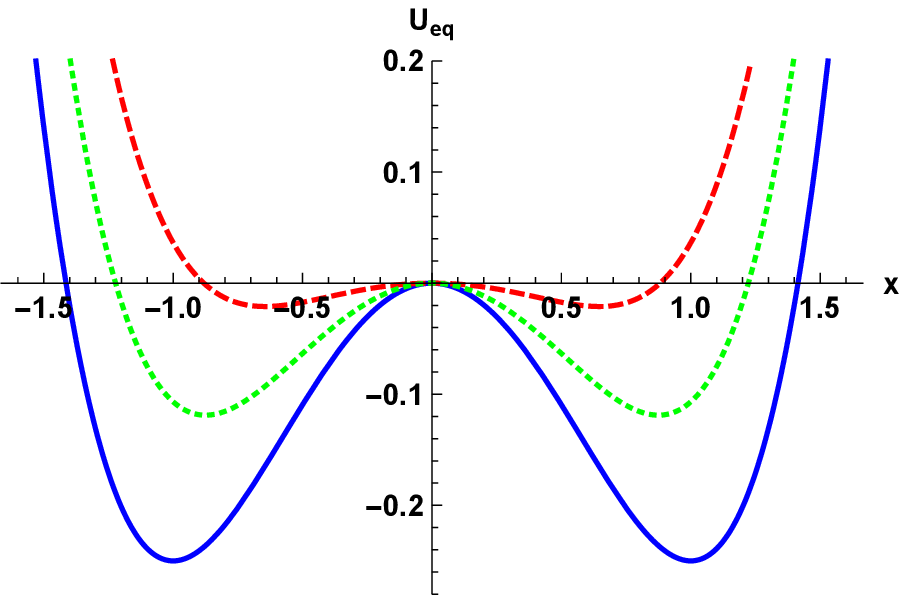}
\includegraphics[height= 4.7 cm]{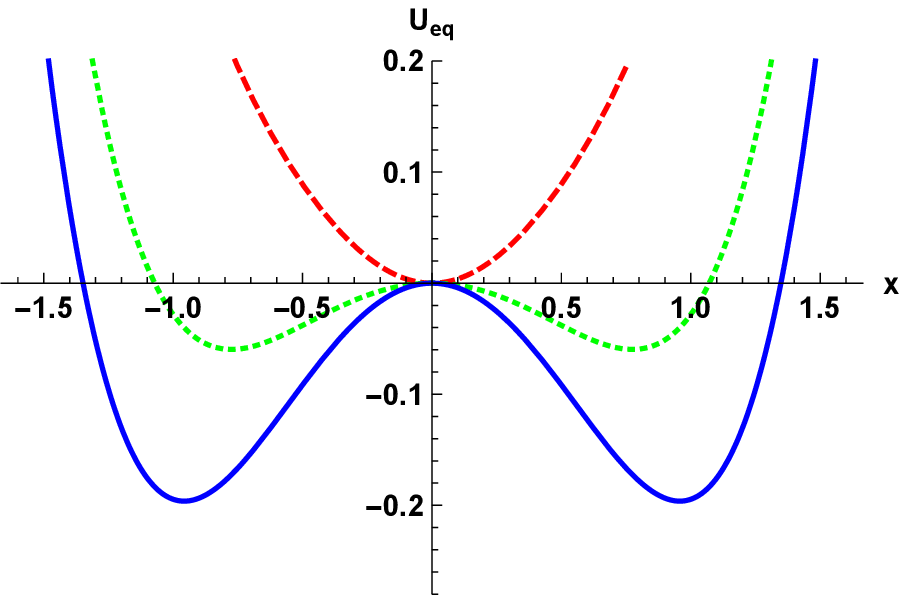}
\caption{Equilibrium potential $U_{\rm eq}(x)$ given by Eq.~(\ref{eq:Ueq}). In (a) we fixed $\sigma=0.5$. The continuous line is plotted 
in the anti-It\^o prescription $\alpha=1$, the dotted line is in the Stratonovich prescription $\alpha=1/2$ and the dashed line corresponds
to the It\^o interpretation $\alpha=0$. In (b), all the curves are computed in the It\^o interpretation. The continuous curve 
is plotted with $\sigma=1/5$, the dotted line with $\sigma=2/5$ and the dashed line with $\sigma=2/3$. In both figures we have fixed $\lambda=1$. }
 \label{fig:Ueq}
 \end{figure}
 
In Figure~\ref{fig:Ueq},  we depict the  equilibrium potential $U_{\rm eq}(x)$ given by Eq.~(\ref{eq:Ueq}) for the simple model specified by 
Eqs.~(\ref{eq:U}) and~(\ref{eq:g}), for different values of the parameters 
$\sigma$ and $\alpha$. In  Figure~\ref{fig:Ueq}-(a), we show the equilibrium potential for $\sigma=0.5$ and different values of the 
stochastic prescription $\alpha=0,1/2,1$. We see that, for $\alpha=1$, $U_{\rm eq}=U$ and the minima are fixed at $x_{\rm min}=\pm 1$. 
However, in the Stratonovich and It\^o prescriptions, the minima are displaced towards the origin.  In Figure~\ref{fig:Ueq}-(b), the three 
curves are computed in the It\^o prescription with different values of the noise $\sigma=1/5, 2/5, 2/3$.   In this case, the minima approach 
zero when the noise grows and, for the value $\sigma=2/3>\sigma_c= 1/2$, the equilibrium potential has only one global minimum at $x_{\rm min}=0$.

\section{Conditional probabilities: path integral representation }
\label{sec:pathintegral}
We are interested in  computing the conditional probability $P(x_f, t_f|x_i, t_i)$  of finding the system in the state $x_f$ at time $t_f$, provided the
system was in the state $x_i$ at a previous time $t_i$. It is useful to express this quantity using a path integral representation~\cite{Miguel2019}.
It can be written as
\begin{equation}
P(x_f, t_f|x_i, t_i)= e^{-\frac{\Delta U_{\rm eq}}{2\sigma^2}} K(x_f, t_f|x_i, t_i)
\label{eq:PK}
\end{equation}
where $\Delta U_{\rm eq}=U_{\rm eq}(x_f)-U_{\rm eq}(x_i)$ and  the \emph{propagator} $K(x_f, t_f|x_i, t_i)$ is given by 
\begin{equation}
K(x_f, t_f|x_i, t_i)=\int [{\cal D}x]\;  e^{-\frac{1}{\sigma^2}\int_{t_i}^{t_f} dt  \; L(x,\dot x)} \; .
\label{eq:Propagator}
\end{equation}
Here, the functional integration measure is
\begin{equation}
[{\cal D}x]={\cal D}x\;{\det}^{-1} g=\lim\limits_{\substack{{N\to\infty} \\ {\Delta t\to 0}}} \prod_{n=0}^N \frac{dx_n}{\sqrt{\Delta t \;g^2(\frac{x_n+x_{n+1}}{2})}}
\label{eq:Dx}
\end{equation}
where $x_0=x_i$  and $x_N=x_f$. 
The Lagrangian can be written in the  form, 
\begin{equation}
 L= \frac{1}{2}\left(\frac{1}{g^2(x)}\right) \dot x^2+V(x) \; ,
\label{eq:L}
\end{equation}
where
\begin{equation}
  V(x) = \frac{g^2}{2}\left[\left(\frac{U'_{\rm eq}}{2}\right)^2 - \sigma^2\left(\frac{U''_{\rm eq}}{2}+\frac{g'}{g} U'_{\rm eq} \right) \right]
  + \frac{\sigma^4}{4}\left(g g'\right)' .
   \label{eq:V}
\end{equation}
The primes mean derivative with respect to $x$. 
Equation~(\ref{eq:Propagator}), with the Lagrangian defined by Eq.~(\ref{eq:L}), correctly describes the dynamics of the Langevin 
Eq.~(\ref{eq:Langevin2}) for arbitrary values of the parameter $0\le\alpha\le 1$~\cite{Miguel2019}. It is important to note that  all the 
information about the stochastic prescription is codified 
in the  structure of the equilibrium potential $U_{\rm eq}(x)$,  contained in the definition of the potential $V(x)$,  Eq.~(\ref{eq:V}).
In this particular representation, the path integral measure given by Eq.~(\ref{eq:Dx}) is discretized symmetrically, allowing us to use 
normal calculus rules in the manipulation of the path integral (for more details on the subtleties of  stochastic calculus in the path 
integral formulation, please see Ref.~\cite{Arenas2012-2}  and references therein).

An interesting observation is  that   Eq.~(\ref{eq:Propagator}) coincides with  the propagator of a quantum particle with position-dependent 
mass $m(x)=1/g^2(x)$ 
moving in a potential $V(x)$, written in the imaginary time path integral formalism $t\to -it$.  The noise $\sigma^2$ plays the role of 
$\hbar$ in the quantum theory.  
At a classical level, the Lagrangian, Eq.~(\ref{eq:L}), represents a particle with variable  mass  moving in a potential $-V(x)$.
The structure of the potential $-V(x)$ (Eq.~(\ref{eq:V})) is much more complex than $U(x)$ or even $U_{\rm eq}(x)$. 

\begin{figure}
\includegraphics[height= 4.7 cm]{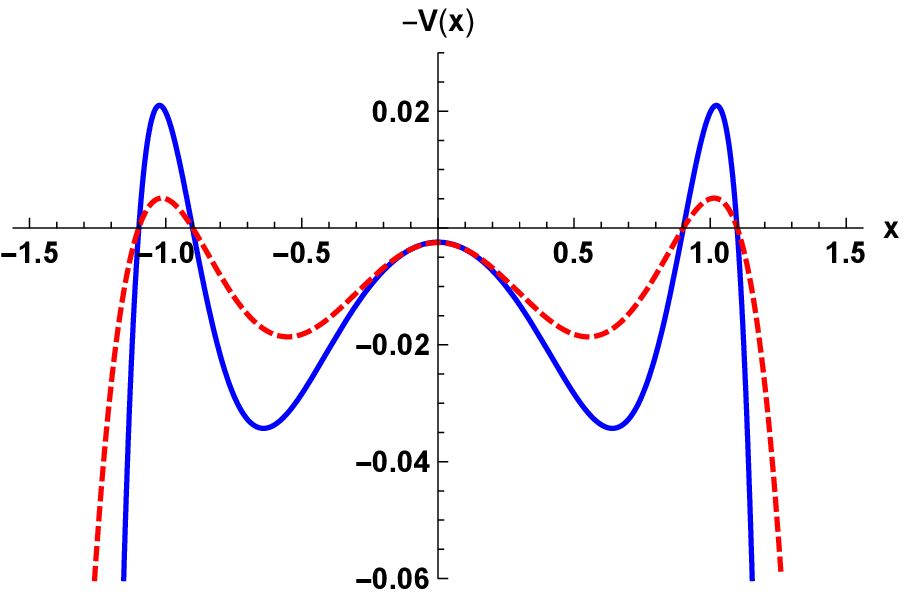}
\centering
\includegraphics[height= 4.7 cm]{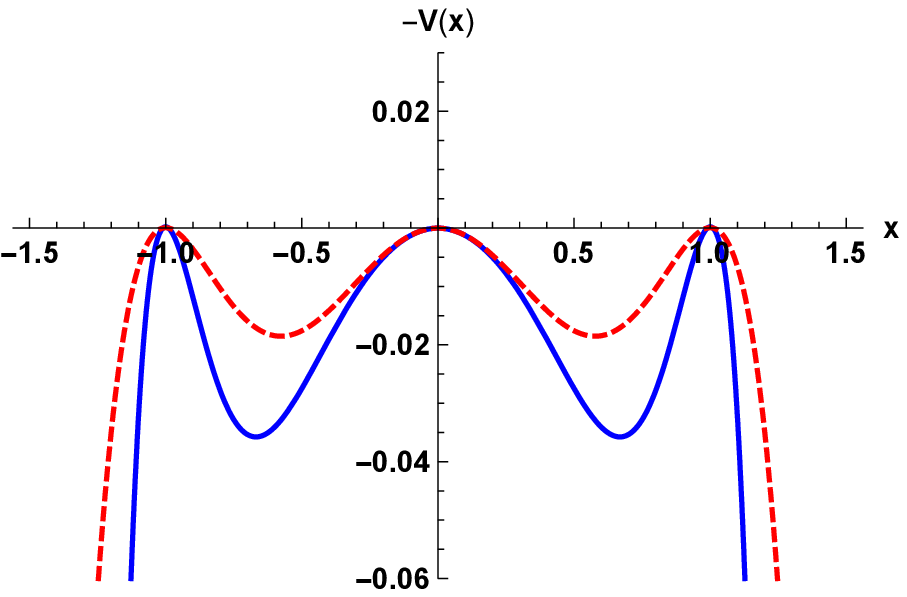}
\caption{Potential $-V(x)$ given by Eq.~(\ref{eq:V}). All the plots are in the It\^o prescription, $\alpha=0$.  The dashed lines 
are the potentials in the additive noise case $g(x)=1$ and the continuous lines correspond with  multiplicative noise, for $g(x)=1+x^2$.  
In (a) we have fixed $\sigma=0.1$ while in (b), $\sigma=0.01$.}
 \label{fig:potentialV}
 \end{figure}

In Figure~\ref{fig:potentialV}, we plotted the potential $-V(x)$ for  the simple model displayed by  Eq.~(\ref{eq:U}).
All the curves have been plotted in the It\^o prescription $\alpha=0$. The dashed lines correspond to the additive noise case $g(x)=1$, 
while the continuous lines represent the potential in the multiplicative noise case, with $g(x)=1+x^2$. In Figure~\ref{fig:potentialV}-(a) we 
fixed $\sigma=0.1$, while in Figure~\ref{fig:potentialV}-(b), $\sigma=0.01$. The first observation is that  $-V(x)$ has three maxima and 
two minima.  The location of both non-zero  maxima roughly coincides with the minima of the potential $U(x)$. The difference is of the 
order of $\sigma^2$.   The main effect of the diffusion function is to increase  the curvature at each maxima with a factor proportional 
to $g^2(x_{\rm max})>1$.  An important feature that will be  relevant to compute conditional probabilities is that the difference between 
the height of the peaks are of the order of $\sigma^2$. Thus, in a weak noise regime, the difference between the three maxima  tends to disappear. 
In the extreme limit of $\sigma\to 0$, the potential $-V(x)$ has three degenerate maxima.  This fact is clearly shown in Figure~\ref{fig:potentialV}-(b). 
It is timely  to note that the structure of $-V(x)$  is quite different from a similar calculus of the tunneling probability amplitude  of a 
quantum particle~\cite{Coleman1979}. In that case, the relevant potential  is $-U(x)$, which has only two maxima. The appearance of a quasi-degenerate 
maximum at $x=0$ is proper of a classical stochastic process, even additive as well as multiplicative.

\section{Fluctuations and time reparametrization}
\label{sec:Weaknoise}

The usual weak noise expansion consists in evaluating the path integral of Eq.~(\ref{eq:Propagator}) in the saddle-point approximation 
plus Gaussian fluctuations. Generally, multiplicative noise induces an integration measure that depends on the diffusion function $g(x)$. 
In Ref.~\cite{Miguel2019}, we have shown how to overcome this problem by means of a time reparametrization. In this section, we briefly review 
this technique since we will use it to compute conditional probabilities. 

The classical equation of motion is
 \begin{equation}
 \frac{d^2 x}{dt^2}=g^2 V' +\frac{g'}{g}  \dot x^2  \;. 
 \label{eq:SadlePoint}
 \end{equation}
Despite the fact that this is a complicated nonlinear equation,  using  time translation symmetry, a first integral can be built up. We have
\begin{equation}
\dot x_{cl}^2=2 g^2_{cl}\left(V_{cl}+H\right) \, .
\label{eq:vsquare}
\end{equation}
Here, $x_{cl}(t)$ is a solution of Eq. (\ref{eq:SadlePoint}). The notation $x_{cl}$ stands for classical solution,  resembling in some sense 
a semiclassical calculation in quantum mechanics. $H$ is an arbitrary constant, $g_{cl}=g(x_{cl}(t))$ and $V_{cl}=V(x_{cl}(t))$.
Then,  the solution of Eq.~(\ref{eq:SadlePoint}) can be expressed by a quadrature,
\begin{equation}
t-t_0= \int_0^{x_{cl}}  \frac{ds}{\sqrt{2  V_{\rm eff}(s)}}\; , 
\label{eq:firstIntegral}
\end{equation} 
where we have defined an effective potential,
\begin{equation}
V_{\rm eff}(x)=  g^2(x)\left[V(x)+H\right].
\label{eq:Veff}
\end{equation}
These expressions have two arbitrary constants, $t_0$ and $H$, that should be determined by means of  the boundary 
conditions $x_{cl}(t_i)=x_i$ and $x_{cl}(t_f)=x_f$. Thus, Eqs.~(\ref{eq:firstIntegral}) and~(\ref{eq:Veff})  implicitly define $x_{cl}(t)$, 
used as a starting point of the weak noise approximation.

Let us assume, for the moment, that, given initial and final conditions, the classical solution $x_{cl}$ is unique. 
 Then, we consider fluctuations around it   
\begin{equation}
x(t)=x_{cl}(t)+\delta x(t) \ , 
\label{eq:xfluct}
\end{equation}
with boundary conditions $\delta x(t_i)=\delta x(t_f)=0$.
Replacing Eq.~(\ref{eq:xfluct}) into Eq.~(\ref{eq:Propagator}) and keeping up to second-order terms in the fluctuations, we find for the propagator
\begin{align}
K(x_f, t_f|& x_i, t_i) =
\label{eq:PropagatorF} \\
& ~~ e^{-\frac{1}{\sigma^2}S_{cl}} \int [{\cal D}\delta x]\; e^{-\frac{1}{2}\int dtdt'\; \delta x(t) O(t,t') \delta x(t') }    \; ,
\nonumber
\end{align}
where the classical action $S_{cl}$ is
\begin{equation}
S_{cl}=\int_{t_i}^{t_f} dt  \; L(x_{cl}(t),\dot x_{cl}(t))
\label{eq:Scl}
\end{equation}
and the fluctuation kernel,
\begin{align}
O(t,t')&=-\frac{d~}{dt}\!\left( \frac{1}{g^2_{cl}}\frac{d\delta(t-t')}{dt}\right)\!+\!\left(\frac{1}{g^2_{cl}}V'_{\rm eff}(x_{cl})\right)'\!\!\delta(t-t').
\nonumber \\
& 
\label{eq:kernel}
\end{align}
In Eq.~(\ref{eq:PropagatorF}),  the functional integration measure is 
\begin{equation}
[{\cal D}\delta x]=\lim\limits_{\substack{{N\to\infty} \\ {\Delta t\to 0}}} \prod_{n=0}^N \frac{d\delta x_n}{\sqrt{\Delta t \;g^2(\frac{x_{cl}(t_n)+x_{cl}(t_{n+1})}{2})}}.
\label{eq:Ddeltax}
\end{equation}

 Due to the time dependence of $g_{cl}=g(x_{cl}(t))$, the fluctuation  kernel $O(t,t')$ is not trivial. On the other hand, the integration measure, 
 Eq.~(\ref{eq:Ddeltax}),  depends on  the diffusion function $g(x(t))$.  
 As a consequence, although the exponent in Eq.~(\ref{eq:PropagatorF}) is quadratic, the evaluation of the functional integral is cumbersome.
 In this case, to compute the fluctuation integral, we make a time reparametrization.
 For concreteness, we introduce a new time variable $\tau$ by means of 
\begin{equation}
\tau=\int_0^t g^2(x_{cl}(t')) dt'\; .
\label{eq:reparametrization}
\end{equation}
This is a nontrivial \emph{local} scale transformation, weighted by the diffusion function evaluated at the classical solution $x_{cl}(t)$. 
Performing this time reparametrization,  the fluctuation kernel transforms as $O(t,t')\to \Sigma(\tau,\tau')$ and takes the simpler form 
\begin{equation}
\Sigma(\tau,\tau')=\left[-\frac{d^2~~}{d\tau^2}+W[x_{cl}]\right]\delta(\tau-\tau')
\label{eq:Sigma}
\end{equation}
where 
\begin{equation}
W(x_{cl})=
\frac{1}{g^2_{cl}}\left(\frac{1}{g^2_{cl}}V'_{\rm eff}(x_{cl})\right)'\; .
\label{eq:W}
\end{equation}
More important, after discretizing the reparametrized time axes $\tau$, the functional integration  measure, Eq.~(\ref{eq:Ddeltax}) becomes  
\begin{equation}
[{\cal D}\delta x]=\lim\limits_{\substack{{N\to\infty} \\ {\Delta \tau\to 0}}} \prod_{n=0}^N \frac{d\delta x_n}{\sqrt{\Delta \tau}} \; ,
\label{eq:Ddeltaxtau}
\end{equation}
in which the function $g(x_{cl})$ has been absorbed in the reparametrization. 

Thus, in the new  time variable $\tau$, the functional integral over fluctuations can be formally evaluated, 
obtaining for the propagator
\begin{equation}
K(x_f, t_f|x_i, t_i)=\left( \det\Sigma(\tau_i,\tau_f)\right)^{-1/2} e^{-\frac{1}{\sigma^2}S_{cl}(t_i,t_f)} \ ,
\label{eq:PropagatorWeaknoise}
\end{equation}
where the relation between $(\tau_i,\tau_f)$ and $(t_i,t_f)$ is given through Eq.~(\ref{eq:reparametrization}). 

Equation~(\ref{eq:PropagatorWeaknoise}) is formally similar to the weak noise expansion in the additive noise case. However, in this case, 
the determinant is written in terms of a rescaled time parameter $\tau$.
Thus, in order to compute a prefactor, we need to  reparametrized the time variable, compute the determinant and, at the end, go back 
to the original time. In Ref.~\cite{Miguel2019} we have successfully used this technique to compute conditional probabilities of an 
harmonic oscillator in a multiplicative noise environment. Here, we will use it to compute conditional probabilities in a double-well set-up. 
 
\section{Probability of remaining in a well}
\label{sec:instanton}
In order to compute conditional probabilities, 
let us consider a potential $-V(x)$  with the general structure displayed in Figure~\ref{fig:potentialV}. 
We will consider that the potential has local maxima at $x=\pm a$ and $x=0$, while it has two minima, at $x=\pm x_p$. 
The difference $|V(a)-V(0)|\sim O(\sigma^2)$, in such a way that  the three maxima are degenerated in the limit $\sigma\to 0$.
As we have mentioned, the maxima at $x=\pm a$, roughly coincide with the minima of the bare potential $U(x)$. The difference is of order 
$\sigma^2$.

We want to compute the probability of remaining in a minimum of $U(x)$, after some time $t$. Let us compute, for instance, 
the probability of remaining in the state $x=-a$, {\em i.e.}, the probability of finding the particle in the state $x=-a$ at 
a time $t/2$, provided it was in the same point, at a time $-t/2$. As the initial and final states coincide, $\Delta U_{\rm eq}=0$ and, 
from Eq.~(\ref{eq:PK}), we see that this conditional  probability  coincides with the propagator, 
$P \left(-a, t/2 | -a,-t/2\right)=K \left(-a,t/2| -a,-t/2\right)$. So, we are interested in the function  $K(-a, -t/2~|-a, t/2)$ for very long 
times, $t\to\infty$.

The main point is that for long times, there are a huge number of solutions (or approximate solutions) of the saddle-point 
equation which need to be considered in order to compute the path integral in the weak noise approximation. 
A trivial solution of Eq.~(\ref{eq:SadlePoint}) with initial and final conditions $x_{cl}(-t/2)=x_{cl}(t/2)=-a$ is simply $x_{cl}=-a$. 
In this case, the multiplicative noise has a trivial effect. Since $x_{cl}$ does not depend on time, the diffusion function $g_{cl}$ is a 
simple constant that renormalizes the noise intensity $\sigma$. Then,  
the contribution of this solution to $K(-a,t/2|-a,-t/2)$ can be easily computed obtaining, 
\begin{equation}
K^{(0)}(-a,t/2|-a,-t/2)= \left(\frac{g^2_aU''_{\rm eq}(a)}{2\pi\sigma^2}\right)^{1/2} \; , 
\label{eq:K0}
\end{equation}
where $g_a=g(a)$. We are using the superscript $(0)$ to indicate the contribution of the constant solution to the propagator.

 \subsection{Instantons/Anti-Instantons}
In the case of potentials with two degenerate maxima, there are topological time-dependent solutions of the equation of motion with finite 
action that interpolate between both maxima. These solutions are called instantons or anti-instantons and should be taken into account 
to compute the propagator. For very large time intervals, well separated superposition of instantons and anti-instantons will also contribute 
to the path integral in a nontrivial way. The technique of summation over these configurations,  usually called instanton/anti-instanton diluted 
gas approximation, was developed by several authors to compute tunneling amplitudes in  quantum mechanics~\cite{Coleman1979,Brezin1977, Bogomonly1980}. 
In stochastic processes, the technique was applied to the case of additive white noise in Ref.~\cite{Caroli1981}, in which the problem of a 
diffusion in a bistable potential was addressed.  Some years later, the same technique was successfully  applied to color noise 
processes~\cite{BrayMcKane1989,McKane1-1990,McKane2-1990,McKane3-1990}. Here, we will apply it to the multiplicative noise case. 
In the rest of this section we will closely follow the calculation of Ref.~\cite{Caroli1981},  emphasizing  those steps that are proper 
of multiplicative noise.

In addition to the constant solution, there are other time-dependent trajectories which begin and end at $x=-a$ for very long time intervals 
that will contribute to the propagator. In our case, the maximum at $x=0$ is quasi-degenerate with $x=\pm a$. For this reason, we  expect 
that trajectories which begin at $x=-a$, go to approximately $x=0$ and then return to the original point, will also have an important weight 
in the functional integral. 
This type of trajectories are not exact solutions of the classical equation of motion, then, there will be a linear term in the fluctuations 
expansion.  However, this term will be $O(\sigma^2)$ since, in the limit $\sigma\to 0$, it should disappear. 

We denote by  $K^{(1)}\left(-a,t/2|-a,-t/2\right)$,  the contribution of the trajectory $-a\to 0\to -a$ to the propagator.
To compute it, we first rewrite the Lagrangian, Eq.~(\ref{eq:L}),  in the following way
\begin{equation}
L= \frac{1}{2}\left(\frac{1}{g^2(x)}\right) \dot x^2+V^{(0)}(x)+ \delta V(x) \ ,
\label{eq:Lmodified}
\end{equation}
where we have defined the quantity
\begin{equation}
\delta V(x)= V(x)-V^{(0)}(x)=\left\{
\begin{array}{ccl}
0,    && x<-x_{p}  \\ 
V_{0}-V_{a}, && x>-x_{p}  
\label{eq:deltaV}
\end{array}\;.
\right.
\end{equation}
In the last expression, $-x_p$ is the position of the minimum of the potential $-V(x)$,  $V_{a}=V(a)=V(-a)$ and $V_{0}=V(0)$.
The specific form of $\delta V(x)$, as well as the specific value $x_p$ are not important. The final results will not depend on such details. 
Thus, the first two terms of Eq.~(\ref{eq:Lmodified})  describe the dynamics of a particle in a potential $-V^{(0)}$ with truly degenerate maxima, 
while  $\delta V(x)\sim O(\sigma^2)$.

Let us compute  asymptotic  solutions of the classical equation of motion  for the potential $-V^{(0)}$.   
We define the ``instanton'',  $x_I(t)$, as the solution with initial and final conditions $x_{cl}(-t/2)=-a$ and $x_{cl}(t/2)=0$, for very large 
values of $t$.  From Eq.~(\ref{eq:firstIntegral}), we have
\begin{equation}
t-t_0= \int^{x_{I}}_{-x_p}\frac{dx}{\sqrt{2g^2(x)(V^{(0)}(x)-V_{a})}} \ ,
\label{eq:instanton}
\end{equation}
where we fixed the conditions $x_I(t_0)=-x_p$ and $H=V_a$.  These parameters guarantee the above-mentioned initial and final conditions. 

We see, from Eq.~(\ref{eq:instanton}), that the integral is dominated by the region in which $V^{(0)}(x)-V_{a}\to 0$. 
It happens for  $x\to 0 >-x_p$ or $x\to  -a< -x_p$. Thus, to compute the integral  we can expand $V^{(0)}(x)$  around $x=0$ and $x=-a$ to 
second order in powers of $x$ and $x+a$, respectively. Thus, in the harmonic approximation we have
\begin{align}
V^{(0)}_{h}(x)&= \left\{
\begin{array}{lcl}
V_a +  \frac{1}{2} V''_{0} x^2,    &&   x>-x_p  \\ 
\\
V_a +  \frac{1}{2} V''_{a} (x+a)^2, &&   x<-x_p
\end{array}
\label{eq:harmonic}
\right. \;.
\end{align} 
Using this approximation, we obtain for the instanton solution 
\begin{align}
x_{I}(t) &\underset{{t\ll t_0}}{\sim} -a+(-x_{p}+a)\; e^{g_a(V''_{a})^{1/2}(t-t_0-\Delta_{ap})} \; ,
\label{eq:instanton-a} 
\\
x_{I}(t)&\underset{{t\gg t_0}}{\sim}\;- x_{p} \;e^{-g_0(V''_{0})^{1/2}(t-t_0-\Delta_{0p})}\;,
\label{eq:instanton-b}
\end{align}
where we  have introduced the finite constants
\begin{align}
\Delta(x_i,&x_j) = \label{eq:inst44} \\
 & \int_{x_i}^{x_j} \frac{dx}{\sqrt{2}}  \left[\frac{1}{g(x)\sqrt{V^{(0)} - V_{a}}} - \frac{1}{g(x_i)\sqrt{V^{(0)}_{h} - V_{a}}}\right],
\nonumber 
\end{align}
in such a way that, in Eq.~(\ref{eq:instanton-b}), $\Delta_{0p}=\Delta(0,x_p)$ and  $\Delta_{ap}=\Delta(a,x_p)$. 

\begin{figure}
\centering
\includegraphics[height= 3.5 cm]{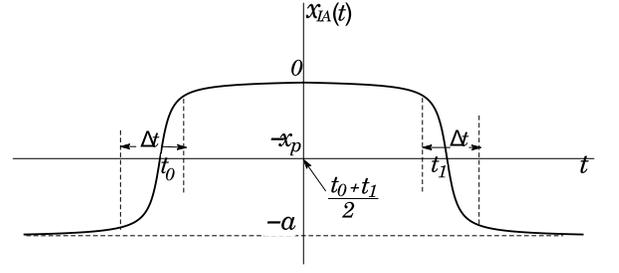}
\caption{Instanton/anti-instanton pair trajectory in the potential $-V^{(0)}(x)$.}
 \label{fig:I-A}
 \end{figure}

The  instanton/anti-instanton pair of trajectories, corresponding with the path $-a\to 0\to -a$, can be written as
\begin{equation}
x_{_{IA}}(t,t_0,t_1)= \left\{
\begin{array}{lcl}
x_{_I}(t-t_0),    &&   t<\frac{t_{0}+t_1}{2}  \\ 
\\
x_{_I}(t_1-t), && t>\frac{t_{0}+t_1}{2}  
\label{eq:I-A}
\end{array}\;,
\right.
\end{equation}
where $x_I(t)$ is given by Eqs.~(\ref{eq:instanton-a}) and~(\ref{eq:instanton-b}).
A typical instanton/anti-instanton trajectory is shown in Figure~\ref{fig:I-A}.
The classical action is  computed by replacing Eq.~(\ref{eq:I-A}) into Eq.~(\ref{eq:Lmodified}) and  integrating in time 
between $t_i=-t/2$ and $t_f=t/2$. 
We find
\begin{align}
S_{IA}(t,t_0,&t_1) = (V_{0}-V_{a})(t_{1}-t_{0}) + V_{a}t  \label{eq:SIA} \\ 
&- \frac{x^{2}_p(V''_{0})^{1/2}}{g_0} e^{g_0(V''_{0})^{1/2}(t_0-t_1+2\Delta_{0p})} \nonumber \\
&+U_{\rm eq}(0) - U_{\rm eq}(a)+ \sigma^2\ln\left|\frac{U''_{\rm eq}(a)\;g^{2}_a\;(x_p+a)}{U''_{\rm eq}(0)\;g_{0}^2 \;x_p}\right| 
\nonumber \\ &+
\frac{\sigma^2}{2}\left[g_{a}^2 U''_{\rm eq}(a)\Delta_{pa} + g_{0}^2 U''_{\rm eq}(0)\Delta_{0p}\right] ,\nonumber  
\end{align}
where we have used the notation $S_{IA}=S_{cl}[x_{IA}]$, {\em i.e.},   the classical action computed at the instanton/anti-instanton 
configuration of Eq.~(\ref{eq:I-A}).
 
The next step is to compute fluctuations around the instanton/anti-instanton solution. 
After the time reparametrization given by Eq.~(\ref{eq:reparametrization}), we are lead to the computation of the determinant  
$\det\hat{\Sigma}(\tau_f,\tau_i)$,  where the operator 
$\hat\Sigma$ is given by Eq.~(\ref{eq:Sigma}),  evaluated at $x_{cl}=x_{IA}(\tau)$. 
Due to time translation invariance, the determinant has zero modes.  Similarly to the original computation of instanton fluctuations~\cite{Coleman1979}, 
we need to properly take into account translation modes, identifying translation fluctuations with the integration over the collective 
variables $t_0$ and $t_1$. We obtain (see Appendix~\ref{Ap:Zeromode}),
\begin{align}
&K^{(1)}\left(-a,\frac{t}{2}\Big\vert -a,-\frac{t}{2}\right) = {\cal N}\int^{t/2}_{-t/2} dt_0 \int^{t/2}_{t_0} dt_1 \;\times
\label{eq:K01t0t1} \\
&  g_a \sqrt{S_I} g_0 \sqrt{S_A}\;
 \left[{\det}'{\hat{\Sigma}\left(\tau_f,\tau_i\right)}\right]^{-1/2}   e^{-\frac{1}{\sigma^2}S_{_{IA}}(t,t_{0},t_{1})}
\nonumber 
\end{align}
where $S_I=S_{cl}[x_I]$, $S_A=S_{cl}[x_A]$ and the prime in the determinant indicates that it should be evaluated excluding the zero modes.  
We use the notation $K^{(1)}$ to indicate the contribution of the path $-a\to 0\to -a$ to the propagator. 
This result is similar to the additive noise case~\cite{Caroli1981}. The main difference is that  the determinant is computed in a 
reparametrized time and the integration over collective variables $t_0$ and $t_1$ are renormalized by the diffusion function.  
The advantage of the reparametrized time is that the operator $\hat \Sigma$ has the simpler form of  Eq.~(\ref{eq:Sigma}) and can be 
computed using the Gelfand-Yaglom theorem~\cite{Dunne2008}. At the end of the calculation, we go back to the original time axes.  
Following  tedious but usual procedures, we finally find 
\begin{equation}
K^{(1)} \left(-a,\frac{t}{2}\Big\vert -a,-\frac{t}{2}\right)= -g_{0}^{2} t \; K^{(0)}  \;\Gamma \ ,
\label{eq:K1}
\end{equation} 
where $K^{(0)}$ is the contribution of the constant solution, given by Eq.~(\ref{eq:K0}), and 
\begin{equation}
\Gamma = \frac{\left(U''_{\rm eq}(a) |U''_{\rm eq}(0)|\right)^{1/2}}{2\pi }
\exp\left\{-\frac{U_{\rm eq}(0) - U_{\rm eq}(a)}{\sigma^2}\right\}\;.
 \label{eq:Gamma}
\end{equation}

We see that the contribution of an instanton/anti-instanton configuration to the propagator at long times, is a linear function of time. 
The structure of the coefficient $\Gamma$ is very interesting.  All the information about the stochastic calculus is hidden in the 
definition of the equilibrium potential, $U_{\rm eq}$.  On the other hand, it does not depend on the details of $U_{\rm eq}(x)$, but instead, 
it depends on the  barrier height, $U_{\rm eq}(a)-U_{\rm eq}(0)$, and on the curvature at each maxima, $U''_{\rm eq}(0)$ and $U''_{\rm eq}(a)$. 
These properties are quite similar with the additive noise case, except for the fact that the original potential $U(x)$ is replaced by the 
equilibrium potential $U_{\rm eq}$  and the time is rescaled by the diffusion function at the maximum of the potential $t\to g^2_0 t$.     In this way, $K^{(1)}$ does not depend on the details of $g(x)$, but only on its 
value at the maxima, $g(0)$ and $g(a)$. 

\begin{figure}
\centering
\includegraphics[height= 3.4 cm]{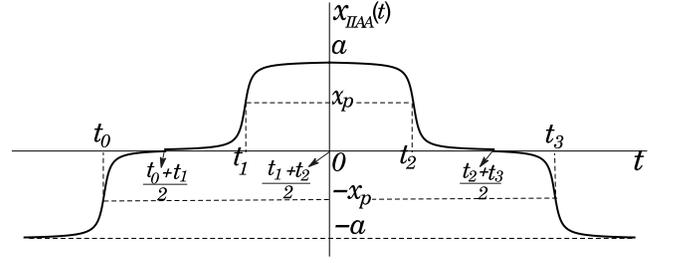}
\caption{Representation of a trajectory of 2-Instanton and 2-Anti-instanton in the potential $-V^{(0)}(x)$.}
 \label{fig:2IA}
 \end{figure}

Due to the structure of the potential $-V(x)$,  there are other trajectories which contribute in a nontrivial way to the propagator; for instance, 
trajectories that begin in $x=-a$, go to $x=a$ passing through $x=0$, and return to $x=-a$. This kind of trajectories contains two instantons and 
two anti-instantons as shown in Figure~\ref{fig:2IA}.  
The contribution of  these trajectories to the propagator can be computed following the same steps of the computation of  the single 
instanton/anti-instanton case.    We find, in this case, 
\begin{align}
&K^{(2)} \left(-a,\frac{t}{2}\Big\vert -a,-\frac{t}{2}\right)= \frac{ (g_0^2 t)^2}{2!} \; K^{(0)}\; \Gamma^2 \ .
\label{eq:K2}
\end{align} 
Thus, trajectories of the type $-a\to  a \to -a$, produce a quadratic time contribution,  the coefficient is simply  
$\Gamma^2$, where $\Gamma$ is given by Eq.~(\ref{eq:Gamma}). 

\subsection{Kramers' escape rate and time reversal transformation}

 To compute the conditional probability of remaining in a minimum after some time $t$, we need to sum up all the 
 trajectories that begin and end at $x=-a$ and which contribute to the propagator in a nontrivial way. Having in mind that $\Delta U_{\rm eq}=0$,
 this probability coincides with the propagator, $P \left(-a, t/2 | -a,-t/2\right)=K \left(-a,t/2| -a,-t/2\right)$. 
  As described above, there are essentially three contributions to these paths:  a constant one, $K^{(0)}$, given by Eq.~(\ref{eq:K0}), 
  a linear term $K^{(1)}$ given by Eq.~(\ref{eq:K1}), corresponding to trajectories $-a\to 0 \to -a$  or, by symmetry, to $a\to 0 \to a$, and,
  finally, a quadratic term $K^{(2)}$ given by Eq.~(\ref{eq:K2}), related to the path $-a \to a \to -a$. 
  
  Consider, for instance, a general 
  trajectory containing $\ell_1$  paths of the type $-a\to 0 \to  -a$ and $\ell_2$ paths of the type $a\to 0 \to  a$, related with the 
  linear function $K^{(1)}$. In addition, we allow $m$ paths of the type $-a\to a\to -a$, related with $K^{(2)}$. Then, this particular 
  trajectory will contribute to the propagator with a term 
\begin{align}
K^{(\ell_1,\ell_2,m)}&\left(-a,\frac{t}{2} \Big\vert -a,-\frac{t}{2}\right) =  \nonumber \\   
& \hspace{1cm} K^{(0)} \frac{(-g_0^2 t)^{\ell_1+\ell_2+2m}}{(\ell_1+\ell_2+2m)!}  \Gamma^{\ell_1+\ell_2+2m} \ .
\end{align}      

By carefully counting the number of different paths which contribute to each trajectory labeled by $(\ell_1,\ell_2,m)$ and summing up, we 
finally arrive at the expression for the conditional probability,  
\begin{align}
P\left(-a,\frac{t}{2}\Big\vert -a,-\frac{t}{2}\right)=\frac{1}{2}K^{(0)} \times \big(1+e^{-t/\tau_k}\big)\; .
\label{eq:P-a-a}
\end{align}

On the other hand, by using the same formalism, we easily find the expression for the conditional probability of finding the system 
in the state $x = a$ at time $t/2$, provided it was in the state $x = -a$ at a previous time $-t/2$,
\begin{align}
P\left(a,\frac{t}{2}\Big\vert -a,-\frac{t}{2}\right)=\frac{1}{2}K^{(0)} \times \big(1-e^{-t/\tau_k}\big)\; .
\label{eq:P+a-a}
\end{align}

In Eqs.~(\ref{eq:P-a-a}) and~(\ref{eq:P+a-a}), the inverse time parameter $\tau_k^{-1}$, which is equivalent to the Kramers' escape rate, is given by 
$\tau_k^{-1}=r_{\rm mult}= g_0^2  \Gamma$. Using Eq.~(\ref{eq:Gamma}), it is explicitly written as
\begin{align}
r_{\rm mult}=g_0^2  \frac{ \sqrt{ U''_{\rm eq}(a)|U''_{\rm eq}(0)|}}{2\pi}\; 
e^{-\frac{\Delta U_{\rm eq}}{\sigma^2}} \ ,
\label{eq:rmult}
\end{align}
with $ \Delta U_{\rm eq}=U_{\rm eq}(0) - U_{\rm eq}(a)$. 

This is one of the main results of our paper.  Comparing Eq.~(\ref{eq:rmult}) with the classical result of Eq.~(\ref{eq:KramersEscapeRate}), 
we clearly see the effect of the multiplicative noise. Notice that the role of the original potential $U(x)$ is now played by the 
equilibrium potential $U_{\rm eq}(x)$ given by Eq.~(\ref{eq:Ueq}). This potential depends not only on the diffusion function $g(x)$ and the noise, 
but also on the stochastic prescription $\alpha$ which defines the original Langevin equation. 
There is also an important  global scaling factor given by $g^2(0)$.  

It is worth to mention that, to the best of our knowledge, there are only few papers where analytic expressions for the escape rate in 
the multiplicative noise case were in fact derived. Indeed, there is no one where different stochastic prescriptions are discussed.  
In Refs.~\onlinecite{Jin2005,FengGuo2011,NingLi2006}, particular examples combining multiplicative with additive noise were treated. 
There seems to be a consensus that in the exponential part of the Arrhenius form, the classical potential should be replaced by an 
effective potential computed from the static solution of the Fokker-Planck equation. However, the values presented for the prefactor 
differ from ours.
As a matter of facts, in all that references, there is no  indication of the discretization prescription used. This fact is quite important in 
multiplicative noise, since different prescriptions correspond to completely different stochastic processes. In such a situation, it is necessary
  to proceed with great care in order to compare analytic expressions and numerical data.     In Ref.~\onlinecite{Rosas2016}, a careful 
treatment of the first time passage was made by focusing on the  Fokker-Planck equation in the Stratonovich prescription. Its result 
coincides with ours for $\alpha=1/2$ in the weak noise limit.  

In order to gain more insight on Eq. (\ref{eq:rmult}), let us compare the Kramers' escape rate with the expression of $r_{\rm mult}$.  Expanding  Eq.~(\ref{eq:rmult}) for weak noise. We obtain
 \begin{equation}
 \frac{r_{\rm mult}}{r_{\rm add}}=|g_0|^{2\alpha} |g_a|^{2(1-\alpha)} \left(1+O(\sigma^2)\right) \; .
\label{eq:raddmult} 
 \end{equation}
 It can be noticed that the relation  between both escape rates does not depend on details of $g(x)$, but on its value at each maxima of 
 $-V(x)$, $x=\pm a$ and $x=0$.  As expected, Eq.~(\ref{eq:raddmult}) depends on the stochastic prescription parameter $\alpha$. For instance, in the case 
 of the Stratonovich prescription, $\alpha=1/2$,   $r_{\rm mult}/r_{\rm add}=g_0 g_a$.  In this case, $g_0$ and $g_a$ have the same weight. 
 On the other hand, in the It\^o interpretation $\alpha=0$,  $r_{\rm mult}/r_{\rm add}=g_a^2$ while, in the thermal prescription, 
 $\alpha=1$, $r_{\rm mult}/r_{\rm add}=g_0^2$.  Indeed,  Eq.~(\ref{eq:raddmult}) is invariant under the transformation
 \begin{align}
   \alpha & \longleftrightarrow  1-\alpha  \\
        0 & \longleftrightarrow    a
 \end{align}
which is nothing but a time reversal transformation~\cite{Arenas2012-2}. The simplest way to understand this symmetry is by noting that 
the instanton solution $x_I(t)$ interpolates between the states $x=-a$ and $x=0$.  The time reversal solution, the anti-instanton  $x_A(t)=x_I(-t)$, makes the 
inverse trajectory, {\em i.e.}, connecting $x=0$ with $x=a$. However, if the forward time process evolves with the $\alpha$ prescription, 
the backward evolution takes place with the $1-\alpha$  prescription. In this sense, one process is the time reversal conjugate 
of the other one. For this reason, the kinetic prescription $\alpha=1$ is also called the anti-It\^o interpretation. In fact, the only time 
reversal invariant prescription is the Stratonovich one, $\alpha=1/2$. For details on the time reversal transformation in multiplicative
noise dynamics, please see Refs.~\cite{arenas2012, Arenas2012-2,Miguel2015}.  

Let us finally mention that the escape rate in the  multiplicative case may be  greater or lower than in the additive case, depending essentially on 
the values of  $g(0)$ and $g(a)$.  Moreover, if the diffusion function $g(x)$ locally  approaches zero at either $x=a$ or $x=0$, 
the escape rate goes to zero.  This effect can be understood from the fact that  the effective curvature of $V(x)$
 approaches zero  and the particle tends to remain in the well for a long time.  
Of course,  our approximation $t \gg \tau_k$ is no longer valid in this limit.     

\section{Numerical simulations}
\label{Ap:Numerics}
 In this section, we perform numerical simulations for the stochastic process driven by the  Langevin  
 equation~(\ref{eq:Langevin2}) with~(\ref{eq:whitenoise}), interpreted in the \emph{generalized Stratonovich} prescription. 
 We use the Euler-Maruyama scheme, which is the simplest algorithm for this task. This algorithm implies an It\^o discretization
 of the stochastic differential equation (SDE). Thus, for a Langevin equation interpreted in a given $\alpha$ prescription,  $0\le\alpha\le 1$,
 it must be transformed to Itô prescription by appropriately changing the drift function $f(x)$.
  As a consequence,  we represent  any $\alpha$ defined SDE  by means of the following It\^o differential equation, 
\begin{equation}
\frac{dx}{dt} =-\frac{1}{2} g^2(x) \frac{dU(x)}{dx}+\sigma^2\alpha g(x)g'(x)  + g(x) \eta(t).
\label{eq:Langevin2-Ito} 
\end{equation}
Eq.~(\ref{eq:Langevin2-Ito}) was obtained from Eq.~(\ref{eq:Langevin2}) by shifting  $f(x)\to f(x)+\sigma^2\alpha g(x)g'(x)$~\cite{Miguel2019}.

\begin{figure}[ht]
\includegraphics[height= 6.5 cm]{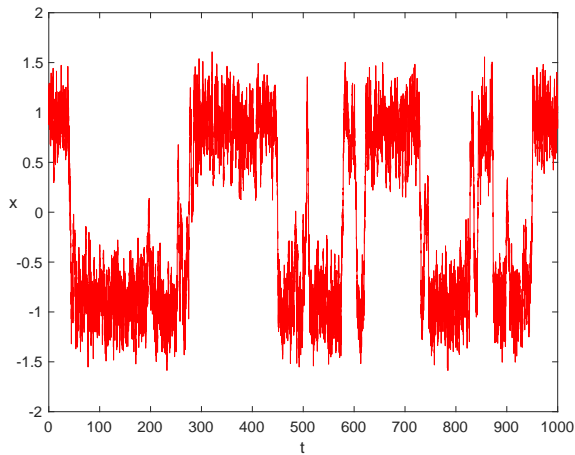}
\caption{$x(t)$,  computed from the integration of Eq.~(\ref{eq:Langevin-numerics}) for a particular realization of the noise, for
$\lambda=0.5$, $\alpha=1/2$ and $\sigma^2=0.095$. Time interval $0<t<1000$   was divided into $2\times 10^4$ steps.}
 \label{fig:noise-realization}
 \end{figure}

 Considering the model given by Eqs.~(\ref{eq:U}) and~(\ref{eq:g}), we explicitly have the It\^o SDE,
\begin{align}
dx = &\frac{x\left(1+\lambda x^2\right)}{2}
\left\{ \left(1-x^2\right)\left(1+\lambda x^2\right)+4 \lambda\sigma^2\alpha  \right\} dt
 \nonumber \\
&+\left(1+\lambda x^2\right)dW \; ,
\label{eq:Langevin-numerics}
\end{align} 
where $W(t)$ is a standard Wiener process with $\langle W(t) \rangle = 0$ and $\langle W(t) W(t') \rangle = \sigma^2{\rm min}(t,t')$.
In Figure~\ref{fig:noise-realization}, we show a typical output for a particular noise realization.
Fixing the initial condition $x(0)=1$, we clearly see the dynamics of the stochastic variable $x(t)$, fluctuating around the 
potential minima $x_{\rm min}\sim \pm 1$, flipping between them  at seemly irregular times.

\begin{figure}
\center
\includegraphics[width = 9 cm,height= 6 cm]{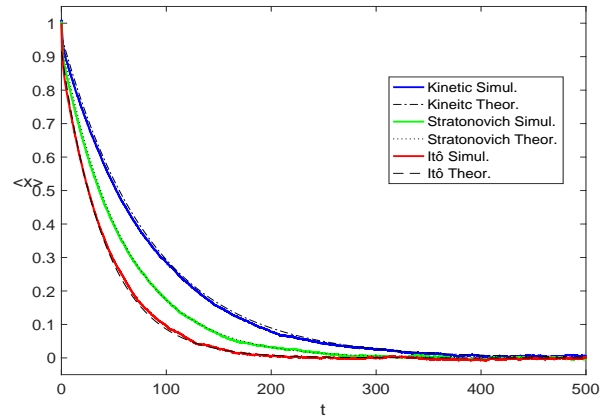}
\caption{$\langle x(t)\rangle$ averaged over $8\times 10^4$ noise realizations.  We fixed the initial condition $x(0)=1$ and the 
parameters  $\lambda=0.5$ and $\sigma^2=0.085$. The three curves corresponds  to three different stochastic prescriptions, $\alpha=0,1/2,1$. 
The continuous lines are the numerical simulations while the 
dashed, dot and dashed-dot lines correspond to a theoretical fitting using Eq.~(\ref{eq:rmult}), in the It\^o, Stratonovich and Kinetics 
stochastic prescription, respectively.}
 \label{fig:meanvalue}
 \end{figure}

 We have computed the mean value $\langle x(t)\rangle$ over different noise realizations. In Figure~\ref{fig:meanvalue}, we show 
 the result of averaging over $8\times 10^4$ configurations of the noise for different values of the stochastic prescription. 
We can observe that, as expected, $\langle x(t)\rangle$ tends to zero exponentially. This means that, at long times, the particle is 
flipping between both potential wells with zero mean value. We can also observe that the typical decay time is not the same  for 
different stochastic prescriptions and, in general, $\tau_I<\tau_S<\tau_K$, where 
$\tau_I$, $\tau_S$ and $\tau_K$ are the decay times in the It\^o, Stratonovich and Kinetic prescriptions. This is consistent with the 
fact observed in Figure~\ref{fig:Ueq}, where we can see that the height  of the equilibrium potential barrier increases with increasing $\alpha$.  

By using the asymptotic conditional probability distributions, Eqs.~(\ref{eq:P-a-a}) and~(\ref{eq:P+a-a}), it is not 
difficult to show that, for $t>>\tau_k$,
\begin{equation}
\langle x(t) \rangle=A \;  e^{-t/\tau_k},
\label{eq:xmean-exp}
\end{equation}  
where $A$ is some constant. We have used Eq.~(\ref{eq:xmean-exp}), with $\tau_k=r_{\rm mult}^{-1}$ computed in Eq.~(\ref{eq:rmult}), 
to compare  the simulations and the theoretical  prediction  in the three cases shown in Figure~\ref{fig:meanvalue}, obtaining  excellent fittings.

\begin{figure}
\includegraphics[height= 6.5 cm]{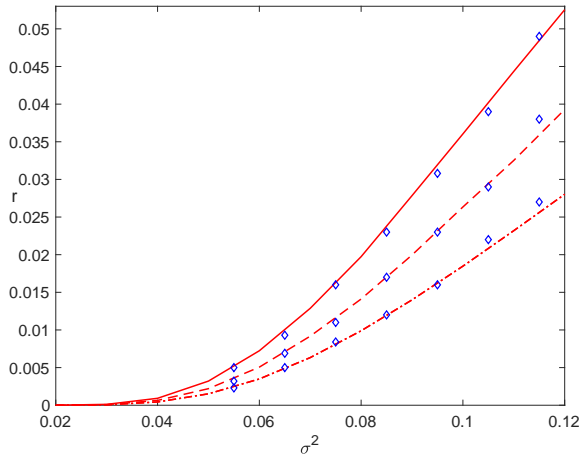}
\caption{Decay rate $r_{mult}$ as a function of the noise intensity $\sigma^2$ computed using Eq.~(\ref{eq:rmult}). 
 The continuous line corresponds to the decay rate in the It\^o prescription. For Stratonovich and kinetic or anti-It\^o interpretation, the 
 decay rate is depicted by the dashed and dot-dashed curves, respectively. The  points (diamonds) were obtained 
 from a linear  fitting of $\ln\langle x(t)\rangle$ through numerical simulations for each case. For all the data, it was fixed $\lambda=0.5$.}
 \label{fig:rate-sigma}
 \end{figure}

In order to have more accurate results, the numerical decay rate $r=\tau^{-1}_k$ can be obtained from  a linear
least-square fitting of $\ln\langle x(t)\rangle$. 
Following this procedure, we studied a wide range of the parameter space $\{ \alpha,\sigma^2\}$ and we compared the 
output with the analytic  decay rate of Eq.~(\ref{eq:rmult}).
In Figure~\ref{fig:rate-sigma}, we show the decay rate $r_{\rm mult}$ as a function of the noise intensity $\sigma^2$ for three 
different values of the stochastic prescription. 
The continuous line represents the decay rate in the It\^o prescription. The Stratonovich interpretation is depicted by the dashed 
line and the dot-dashed curve shows the decay rate in the Kinetic or anti-It\^o prescription. The diamonds are numerical results obtained 
by the least-square fitting of $\ln\langle x(t)\rangle$ in each case.  We can observe an excellent agreement over almost all 
the noise range. As expected, there is a small deviation for larger values of the noise, since in these 
cases $\Delta U_{\rm eq}/\sigma^2\gtrsim 1$, and the Arrhenius form is no longer a good approximation.  

\begin{figure}
\includegraphics[height= 7.5 cm]{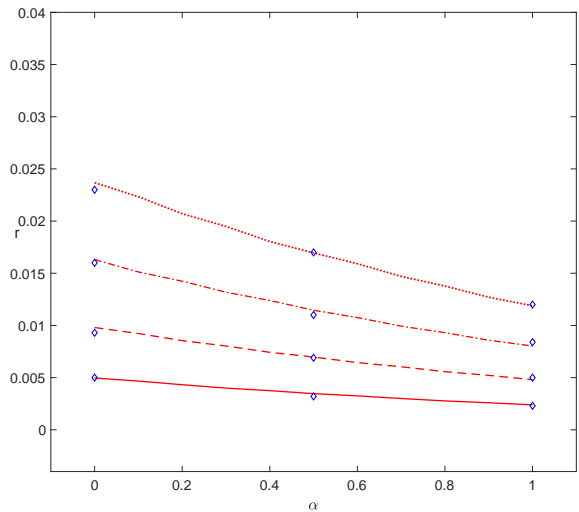}
\caption{Decay rate $r=\tau_k^{-1}$ as a function of the stochastic prescription $\alpha$ obtained from Eq.~(\ref{eq:rmult}) for different values
of $\sigma^2$. The continuous line is plotted for $\sigma^2=0.055$, dashed line corresponds to $\sigma^2=0.065$, 
while the dot-dashed and dotted lines correspond to $\sigma^2 = 0.075$ and $\sigma^2 = 0.085$, respectively.  The points (diamonds) results from
numerical simulation, computed by linear fittings of $\ln\langle x(t)\rangle$. Parameter $\lambda=0.5$ was fixed for all the curves.}
 \label{fig:rate-alpha}
 \end{figure}

In Figure~\ref{fig:rate-alpha} we show the decay rate $r_{\rm mult}$ as a function of the stochastic prescription 
$0\le\alpha\le 1$, for different values of the noise from $\sigma^2=0.055$ to $\sigma^2=0.085$. 
We observe an excellent agreement between the theoretical predictions and the data computed from the numerical simulation of the Langevin equation.  
In this figure, the continuous line  was plotted  fixing $\sigma^2=0.055$ and has a perfect match with the numerical results. We expect that 
lower values of the noise produce still better results. However, for these values, the time decays are huge, 
being on the order of $t = 1000$ for $\sigma^2=0.055$.
So, in order to have statistics for a lower noise range, it would be necessary to simulate for very longer times scales considering a big number of 
noise realizations. Of course, this consumes much more computational resources.

\section{Summary and conclusions}
\label{sec:discussion}
We have considered the problem of a particle in a symmetric double-well potential $U(x)$, with a dynamics driven by an overdamped 
multiplicative Langevin equation characterized by a symmetric diffusion function $g(x)=g(-x)$. The stochastic differential 
equation was defined in the  {\em generalized Stratonovich} prescription, parametrized by a continuum parameter  $0\le\alpha\le 1$. This prescription 
contains the usual stochastic interpretations for particular values of the parameter $\alpha$.  
Indeed,  $\alpha=0,1/2,1$ corresponds to  the usual It\^o, Stratonovich and Kinetic prescriptions, respectively.

We have provided a path integral technique 
to compute conditional probabilities in the weak noise approximation for arbitrary values of the parameter $\alpha$.  
Interestingly, all the dependence of $\alpha$ is codified in the equilibrium potential $U_{\rm eq}(x)$, obtained by means of a static 
solution of the associated Fokker-Planck equation.

It was introduced a local time reparametrization, which allows to 
exactly integrate fluctuations around saddle-point solutions.  
Conditional probabilities  were computed for long time intervals  by  generalizing the 
instanton/anti-instanton diluted gas approximation, already developed for the additive noise case~\cite{Caroli1981}. 
From these probabilities, the escape rate was computed in the same approximation and the result was compared with the Kramers' escape rate 
 for additive noise dynamics.  
 
  The main result of the paper is given by  Eq.~(\ref{eq:rmult}). 
We found that the general structure of the escape rate keeps the Arrhenius form of the Kramers' result. The main corrections are  twofold. 
First, the equilibrium potential $U_{\rm eq}(x)$ of Eq.~(\ref{eq:Ueq}) plays the role of the bare potential $U(x)$.  
The potential $U_{\rm eq}(x)$  is generally different from $U(x)$ in the multiplicative noise case, depending on the diffusion 
function and the stochastic prescription $\alpha$. Indeed, the only 
stochastic prescription in which $U_{\rm eq}(x)=U(x)$ is the anti-It\^o prescription $\alpha=1$.   Moreover, there is a global 
scale factor $g^2(0)$ that has its origin in the time reparametrization necessary to correctly  compute fluctuations.  

In the weak noise limit, we found a simple relation between the Kramers' escape rates computed with additive and multiplicative noise,  given by  
Eq.~(\ref{eq:raddmult}). The obvious consistency check is that  $r_{\rm mult}/r_{\rm add}=1$ in the limit $g(x)\to 1$ (or $\lambda\to 0$ in 
the particular example). 
In addition, we observe that $g(0)$ and $g(a)$ enter with different weights depending on the prescription parameter $\alpha$. These weights are 
consistent with a time reversal transformation, which relates a stochastic process in the $\alpha$ prescription with its time reversal 
conjugate $1-\alpha$. Indeed, the Stratonovich convention $\alpha=1/2$ is the only one with time reversal invariance and, in this case, both maxima 
enter with the same weight.    

Finally, we have made extensive Langevin simulations to test the accuracy of our expressions. We have explored a huge region of the 
parameter space $\{\sigma,\alpha\}$, in which the high barrier approximation, $\Delta U_{\rm eq}/\sigma^2>>1$, is well defined. 
We have found a very good agreement for all values of the stochastic prescription.

Although we have presented results for a system with full reflection symmetry $x\to -x$, the methods developed in this paper are completely 
general. We hope to communicate results for  a more general non-symmetric case in the near future. Moreover, having analytic expressions 
for the conditional probability we can face the problem of  stochastic resonance in multiplicative noise processes in a more solid bases.  

\acknowledgments
The Brazilian agencies, {\em Funda\c c\~ao de Amparo \`a Pesquisa do Rio
de Janeiro} (FAPERJ), {\em Conselho Nacional de Desenvolvimento Cient\'\i
fico e Tecnol\'ogico} (CNPq) and {\em Coordena\c c\~ao  de Aperfei\c coamento de Pessoal de N\'\i vel Superior}  (CAPES) - Finance Code 001,  
are acknowledged  for partial financial support. MVM is partially supported by a Post-Doctoral fellowship by CNPq.

\appendix
\section{Zero modes in the multiplicative case}
\label{Ap:Zeromode}
The relation of zero modes of the fluctuation operator and translation invariance is very well known in quantum mechanics~\cite{Coleman1979}, 
as well as in additive noise stochastic dynamics~\cite{Caroli1981}.  In this appendix, we focus on the effect 
produced by the diffusion function $g(x)$ in a multiplicative noise stochastic system. 

Let us consider the instanton function $x_I(t)$ as a solution of the equation of motion Eq.~(\ref{eq:SadlePoint}), with boundary 
conditions $\lim_{t\to -\infty} x_I(t)=-a$ and $\lim_{t\to \infty} x_I(t)=0$,  where 
$-a$ and $0$ are the positions of a minimum and the local maximun of $U_{\rm eq}(x)$, respectively. In the weak noise approximation, 
these values coincide with two local maxima of $-V(x)$ as shown in Figure~\ref{fig:potentialV}.
It is not difficult to show that $dx_I/dt$ is a zero mode of the fluctuation operator Eq.~(\ref{eq:kernel}).
To see this, we consider
\begin{align}
\int dt' O(t,t')& \frac{dx_I(t')}{dt'}= \\ 
&-\frac{d}{dt}\left(\frac{1}{g^2} \frac{d^2 x_I}{dt^2}\right)+\left(\frac{1}{g^2}V'_{\rm eff}\right)'\frac{dx_I}{dt} = 
\nonumber \\
&-\frac{d}{dt}\left(\frac{1}{g^2} V'_{\rm eff}\right)+\left(\frac{1}{g^2}V'_{\rm eff}\right)'\frac{dx_I}{dt} =0
\nonumber
\end{align}
where in the first term of the last line we have used $d^2x_I/dt^2=V_{\rm eff}$ and in the second term we used the chain rule. 

Thus, the fluctuation operator has a normalized zero mode of the form
\begin{equation}
\eta_0(t)= A \frac{d x_I(t)}{dt}\ ,
\end{equation}
where $A$ is a normalization constant. To determine it, we impose, 
\begin{align}
\int dt\;   \eta_0^2(t)= A^2 \int dt \; \left(\frac{d x_I}{dt}\right)^2=1
\end{align}
and, thus, the normalization constant reads 
\begin{equation}
 A^{-2}= \int dt \; \left(\frac{d x_I}{dt}\right)^2  .
\label{app:A}
\end{equation}

The action computed at the instanton solution is 
\begin{equation}
S_I=\int dt\left\{ \frac{1}{2g^2(x_I)} \left(\frac{dx_I}{dt}\right)^2+V(x_I)\right\}.
\end{equation}
Using the equations of motion, it can be written as 
\begin{equation}
S_I=\int dt\;  \frac{1}{g^2(x_I)} \left(\frac{dx_I}{dt}\right)^2.
\end{equation}
Since the zero mode has a small support around $t_0$, in the thin-wall approximation we can write with good accuracy
\begin{equation}
S_I\sim  \frac{1}{g^2_a}\int dt \left(\frac{dx_I}{dt}\right)^2,
\end{equation}
where $g_a=g(a)$. Replacing this result in Eq.~(\ref{app:A}) we finally find the normalized zero mode
\begin{equation}
\eta_0(t)= \frac{1}{g_a\sqrt{S_I}} \frac{d x_I(t)}{dt}.
\label{app:eta0}
\end{equation}

In order to compute fluctuations, we perform a local time reparametrization given by  Eq.~(\ref{eq:reparametrization}). 
We are lead to 
the computation of the integral
\begin{equation}
I_F=\int [{\cal D}\delta x] \; e^{-\frac{1}{2}\int d\tau \delta x(\tau) \left(-\frac{d^2~}{d\tau^2}+W[x_{cl}]\right)\delta x(\tau)}\ ,
\end{equation}
where $W$ is given by Eq.~(\ref{eq:W}). 
To compute it, we expand fluctuations in eigenfunctions of the fluctuation operator, taking special care with the translational modes that 
are responsible for the zero mode. 
We write the fluctuation field in the following form
\begin{equation}
\delta x(\tau)= c_0 \psi_0(\tau-\tau_0)+\sum_{k=1}^\infty c_k \psi_k(\tau-\tau_0) ,
\end{equation}
where $\psi_k$ are eigenvectors 
\begin{equation}
\left(-\frac{d^2~}{d\tau^2}+W[x_{cl}]\right)\psi_n(\tau)=\lambda_n\psi_n(\tau)
\label{app:eigenfunction}
\end{equation}
with eigenvalues $\lambda_k\neq 0$ and the zero mode  in the reparametrized variable reads
\begin{equation}
\psi_0(\tau)= \frac{1}{g_a\sqrt{S_I}} \; g^2(x_I(\tau)) \frac{d x_I(\tau)}{d\tau}\; .
\label{app:psi0}
\end{equation}

The functional measure can be written in terms of the coefficients $c_k$ as
\begin{equation}
{\cal D}\delta x=dc_0 \prod_{k\neq 0} dc_k \ .
\end{equation}
Computing the variation of fluctuations under time translation, we have that 
\begin{equation}
d\delta x(\tau)= \frac{d x_I}{d\tau} d\tau_0\ .
\label{app:dxt0}
\end{equation}
On the other hand, a variation in the zero mode reads
\begin{equation}
d\delta x(\tau)= \frac{1}{g_a\sqrt{S_I}} \; g^2(x_I(\tau)) \frac{d x_I}{d\tau}  dc_0 \ .
\label{app:dxc0}
\end{equation}

Now, comparing Eqs.~(\ref{app:dxt0}) and~(\ref{app:dxc0}) and 
using the reparametrization identity $d\tau/dt=g^2(x_I)$, we immediately find
\begin{equation}
dc_0=g_a \sqrt{S_I} dt_0 \ .
\end{equation}

In this way, 
\begin{align}
I_F&=\int g_a \sqrt{S_I} dt_0\int \left(\prod_{k\neq 0} dc_k\right) \exp\left(-\frac{1}{2}\sum_n \lambda_n(\tau_0) c_n^2\right) 
\nonumber \\
&=\int g_a \sqrt{S_I} dt_0
\left( \prod_{k\neq 0} \lambda_k^{-1/2}(\tau_0)\right)\nonumber \\
&=\int   dt_0 \;g_a\sqrt{S_I}\; \left( {\det}'\left(-\frac{d^2~}{d\tau^2}+W[x_{cl}]\right)\right)^{-1/2}
\end{align}
where the prime means that the determinant should be computed without the zero mode. 

Thus, the usual interpretation of the zero mode as an integration in the collective variable $dt_0$ is still valid in 
the multiplicative case. However, the constant of proportionality is renormalized by the diffusion function $g_a$, computed at 
the minimum of the potential.

The same reasoning applies to the anti-instanton solutions. However, in this case, the variation is proportional to $g_0 \sqrt{S_A} dt_1$,
where $g_0$ is evaluated at the maximum of the potential and $S_A$ is the classical action evaluated at the anti-instanton solution.
This analysis leads to Eq.~(\ref{eq:K01t0t1}) for $K^{(1)}$. 

%

\end{document}